\documentclass[aps,prl,amssymb,twocolumn,showpacs]{revtex4}

\usepackage{amsmath,amssymb}
\usepackage{graphicx}
\def\ca{c_\textsc{A}}
\def\cm{c_\textsc{H}}
\def\rah{r_{\textsc{AH}}}

\begin{document}
\title{Internal Motility in Stiffening Actin-Myosin Networks}

\author{J\"org Uhde, Manfred Keller, Erich Sackmann}
\affiliation{Physik Department, Technische Universit\"at M\"unchen,
  D-85747 Garching, Germany}

\author{Andrea Parmeggiani$^1$, Erwin Frey$^{1,2}$}
\affiliation{$^1$Hahn-Meitner-Institut, Abteilung Theorie, Glienicker
  Strasse 100, D-14109 Berlin, Germany \\ $^2$Fachbereich Physik,
  Freie Universit\"at Berlin, Arnimallee 14, D-14195 Berlin, Germany}

%\pacs{ 87.15.Ya biomolecules : Fluctuations, 87.15.La Mechanical
%properties, 87.16Ka Filaments, microtubules, their networks, and
%supramolecular assemblies}

%87.15.Kg Molecular interactions; membrane-protein interactions
%87.15.Nn Properties of solutions; aggregation and crystallization
%of macromolecules
%87.16.Nn Motor proteins (myosin, kinesin, dynein)


\date{\today}

\begin{abstract}
  We present a study on filamentous actin solutions containing heavy
  meromyosin subfragments of myosin~II motor molecules. We focus on the
  viscoelastic phase behavior and internal dynamics of such networks
  during ATP depletion.  Upon simultaneously using micro-rheology and
  fluorescence microscopy as complementary experimental tools, we find
  a sol-gel transition accompanied by a sudden onset of directed
  filament motion. We interpret the sol-gel transition in terms of
  myosin~II enzymology, and suggest a ``zipping'' mechanism to explain
  the filament motion in the vicinity of the sol-gel transition.
\end{abstract}

\maketitle

Eucaryotic cells show an amazing versatility in their mechanical
properties. Not only can they sustain stresses ranging from some
tenths to hundreds of Pascals, but they can equally well perform such
complex processes as cytokinesis and cell locomotion. A vital role for
these and other cellular functions is played by the cytoskeleton, the
structural framework of the cell composed of a network of protein
filaments. A major component is filamentous actin (F-Actin), whose
physical properties are by now well
characterized~\cite{legoff-etal:02}.  The length distribution, spatial
arrangement and connectivity of these filaments is controlled by a
great variety of regulatory proteins.

An important family of these regulatory proteins are
cross-linkers, which can be further classified as {\em passive} or
{\em active}. The function of {\em passive cross-linkers}, e.g.
$\alpha$-actinin, is mainly determined by their molecular
structure and the on-off kinetics of their binding sites to actin.
Upon changing the association-dissociation equilibrium and hence
the degree of cross-linking by varying the temperature the network
can be driven from a sol into a gel state~\cite{tempel-etal:96}.
Depending on both the concentration and the affinity of these
cross-linkers for F-actin there is a tendency to either form
random networks or bundles~\cite{wachsstock-schwarz-pollard:93}.
Motor proteins of the myosin family can also act as {\em active
cross-linkers}. When both of its functional head groups are bound
to two different filaments they can use the energy of
adenosine-triphosphate (ATP) hydrolysis to exert relative forces
and motion between them. However, such an event is very unlikely
under physiological conditions and ATP saturation, because then
myosin~II spends only a short fraction of its chemomechanical
cycle attached to the filament (duty ratio: $r \lesssim
0.02$~\cite{howard_book}). Active relative transport yet becomes
possible due to the concerted action of several motors if in vitro
myosin~II proteins assemble into multimeric
minifilaments~\cite{humphrey-etal:02}.

In this letter we study actin networks containing the heavy meromyosin
(HMM) subfragment lacking the light meromyosin domain responsible for
myosin~II assembly. Our focus is on the viscoelastic phase behavior
and internal dynamics of such networks during ATP depletion.  We use
an experimental setup combining microrheology with fluorescent
microscopy of labeled filaments. This allows us to identify a sol-gel
transition accompanied by a sudden onset of directed filament motion.

Monomeric actin (G-actin) was prepared from rabbit skeletal muscle
as described by Spudich and Watt \cite{Spudich:1971}. In order to
reduce residual cross-linking and capping proteins it was purified
by an additional step using gel column chromatography
(Sephacryl~\mbox{S-300}). Actin was sterile filtered and kept in
G-buffer \cite{Spudich:1971} on ice. Myosin~II was obtained from
rabbit skeletal muscle according to a procedure by Margossian and
Lowey \cite{Margossian:1982}. Then, HMM was prepared from this
myosin~II by TLCK-treated chymotrypsin digestion following Kron
\textit{et al.}\ \cite{Kron:1991} with one modification: After the
chymotrypsin treatment the solution was dialyzed against a large
volume of low-salt buffer. The concentration was determined by
absorption spectroscopy (absorption coefficient $= 0.64$~cm$^2$/mg
at $280$~nm) \cite{Weeds:1977}.  HMM was stored either by rapid
immersion of droplets in liquid nitrogen or by mixing with $30$~\%
sucrose. The frozen droplets or small aliquots of the mixed
solution where stored at $-70~^{\circ}$C.  Function tests for both
myosin and HMM were performed by motility assays according to Kron
and Spudich \cite{Kron:1986}.

Fluorescently labelled reporter filaments were produced by
polymerizing G-actin in slightly modified F-buffer
(\cite{Spudich:1971}, using $0.2$~mM ascorbic acid instead of DTT)
in presence of equimolar
tetra\-methyl\-rhodamine\-iso\-thio\-cyanate labelled phalloidin
at room temperature. The actin concentration was $5$~$\mu$M.  For
storage, the reporter filaments were kept on ice and used within
three days.  The surrounding actin network was prepared in
F-buffer~\cite{Spudich:1971}. The ATP concentration was adjusted
to values between $150$~$\mu$M and $1$~mM. An antioxidant system
($0.2$~mg/mL glucose, $0.05$~mg/mL glucose oxidase, $0.01$~mg/mL
catalase, $0.5$ Vol.-\% mercaptoethanol) was added to the buffer
solution. Then, approximately $1.5$--$3.0$~$\mu$L of the
fluorescently labelled reporter filaments were carefully diluted
in $0.5$~mL of this antioxidant buffer.  Second to last, HMM was
added. With the final addition of G-actin polymerization was
started.  We used $200$~$\mu$L of this solution for fluorescence
microscopy measurements. The rest was carefully mixed with
magnetic beads (Dynabeads~M-450, Dynal, Hamburg, Germany;
4.5~$\mu$m radius) and used for magnetic tweezers experiments, as
described in~\cite{Keller:2001}.

All experiments were performed with freshly prepared actin kept on ice
for at most $10$ days. We worked with $9.5$ and $19$~$\mu$M actin
solutions, corresponding to an actin concentration $\ca$ of $0.4$ and
$0.8$~mg/mL, equivalent to a mesh size $\xi\sim0.6$ and $\sim
0.4~\mu$m, respectively, \cite{Schmidt:1989}. The ratio of labelled to
unlabelled actin was between $1:300$ and $1:1000$. The molar ratios of
actin to HMM were $\rah= 8, \, 12, \, 25, \, 50$, and $175$,
corresponding to HMM concentrations between $360$ and $55$~nM.  After
approximately $15$--$20$~min, $90$~\% of G-actin is polymerized into
F-actin~\cite{Detmers:1981}. Measurements were started at least
$5$~min after the polymerization was initiated. $80$\% of the reported
filament transport events happened $15$~min or later after
polymerization start.

Fluorescence measurements were performed in a closed sample chamber
(containing approximately $100$~$\mu$L) with an inverse microscope
(Zeiss Axiovert 200, $100\times$ oil immersion objective (N.A.~1.3))
and a CCD camera (Orca C4742-95-12ER, Hamamatsu, Herrsching, Germany).
$12$~bit images of size $672\!  \times \! 512$ pixels (using $2 \!
\times \! 2$ binning) were stored directly on the hard disk of a
personal computer using the on-line image processing software ``Open
Box'' developed in this laboratory~\cite{Keller:2001}.  The frame rate
was approximately $17$~Hz.  A given region in the sample could be
observed for several minutes before the contrast of the image was
decreased by photo bleaching significantly.

\begin{figure}[htb]
  \includegraphics[width=\columnwidth]{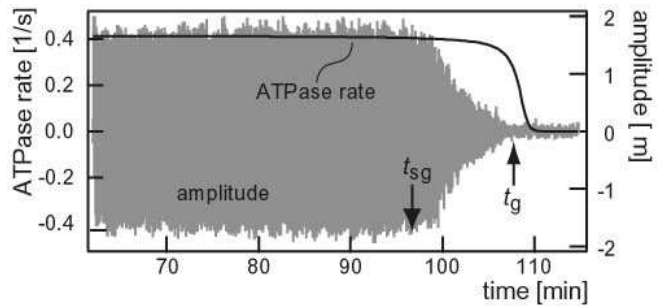}
\caption{\label{fig:SGT} Amplitude of an oscillating magnetic
  bead embedded in an actin-HMM network showing an abrupt transition
  from a non-cross-linked F-actin solution to a gel state after
  $105$~min. The initial ATP concentration was $145$~$\mu$M, $\ca=
  9.5$~$\mu$M , $\rah=175$, $f=2.55$~pN and the oscillation frequency
  $150$~mHz.  The solid line shows the time dependence of the ATPase
  rate (see text).}
\end{figure}

We now discuss our results obtained with a magnetic tweezers
micro-rheometer~\cite{Keller:2001,Schmidt:1996}.  Magnetic beads
of $4.5$~$\mu$m diameter, embedded in the network during
polymerization, were subjected to an oscillating magnetic force of
$f=1.5$--$3$~pN. Figure \ref{fig:SGT} shows the resulting
oscillatory motion, analyzed by a particle tracking procedure with
an accuracy better than $10$~nm and time resolution less than
$8$~ms.
All our data show a characteristic time $t_\text{sg}$, where the bead
amplitude starts to shrink drastically. This is followed by a
transition regime, $t_\text{sg} \leq t \leq t_\text{g}$, during which
the bead amplitude falls below the noise level. We interpret this as a
sol-gel transition caused by the continuous depletion of ATP. For
$t\leq t_\text{sg}$, ATP is present in abundance producing a fast
release of HMM from actin such that it cannot act as a cross-linker.
In this first phase we have a network with the mechanical properties
of a semidilute solution of F-actin. When ATP starts to lack, the
residence time of HMM on actin increases, which in turn induces
progressive cross-linking of the filaments leading to a stiffening of
the network. In addition to their role as passive cross-linkers HMM
also can actively generate forces between filaments. These two
different roles are not easily distinguished when looking at rheology
data alone, but become evident from the dynamics of the embedded
reporter filaments, which we discuss below.

Upon using known features of myosin~II enzymology we can give an
almost quantitative picture of the dynamics of the HMM molecules in
the actin network and the resulting time-dependence of the bead
amplitudes during the process of ATP consumption. A rough estimate of
the consumption rate is $k_c \simeq [\text{ATP}]_0/[\text{M}]_0
t_\text{g}$, where $[\text{ATP}]_0$ is the initial ATP concentration
and $[\text{M}]_0$ is the total HMM concentration.  This estimate
neglects cooperative effects of preferential binding of HMM molecules
close to already bound HMMs~\cite{orlova97}.  From the experimental
data in Fig.\ \ref{fig:SGT} one gets $k_c \approx 0.4~$s$^{-1}$ for
each HMM, which is much closer to the ATPase rate in the absence of
actin than in a saturated F-actin solution~\cite{howard_book}.  We
explain this as a consequence of the diffusive search of the HMM for
an actin filament being the rate limiting process. The characteristic
rate of search and binding to an actin filament can be estimated as
$k_s \sim k_d \phi$, where $k_d\simeq 6 D \xi^{-2}$ is the rate for
diffusing a distance comparable to the mesh size $\xi$; $D$ is the
diffusion constant of HMM in water and $\phi$ is the volume fraction
of actin in solution. From experimental data we estimate~\footnote{For
  the estimate we take the Einstein-Stokes relation $D=k_B T/ 6\pi
  \eta_w r$, where $\eta_w = 10^{-3}~$ kg s/m, $k_\text{B} T = 4 \cdot
  10^{-21}$~J, and the hydrodynamic radius of HMM is $r \sim 20$~nm.
  Our estimate for the volume fraction is $\phi \sim 0.1 \%$ within an
  uncertainty of $30\%$~\cite{keller:thesis}.}  $k_s \sim 0.3$
s$^{-1}$ which is of the right order of magnitude.

Starting from these considerations we calculated the ATP consumption
rate as a function of time using known rates from the literature
\footnote{We use the enzymatic rates of table 14.2 in
  Ref.~\cite{howard_book}.  A detailed analysis and study of the
  inhibition effect of ADP will be published in a more extended
  article~\cite{keller:unpublished}.} with the exception of $k_s$,
which was computed along the lines explained above. The results are
shown as the solid line in Fig.\ \ref{fig:SGT}. The sol-gel transition
occurs at a time $t_\text{sg}$ when ATP depletion becomes significant
and the typical ATPase rate starts to decrease strongly.  Note that we
have taken into account the inhibition effect of ADP, which increases
in solution because of the ATP hydrolysis. Indeed, ADP plays an
important role in both the shape of the ATPase rate and in the
estimate of $t_\text{sg}$ and $t_\text{g}$. For example, an increase
of the initial ADP concentration decreases $t_\text{sg}$ while it
increases $t_\text{g}$~\cite{keller:unpublished}.

\begin{figure}[htb]
  \includegraphics[width=\columnwidth]{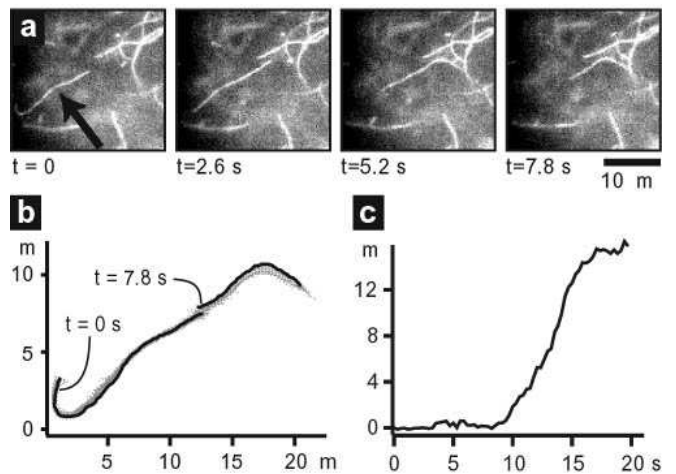}
\caption{\label{fig:BetterSequence} Sequence of fluorescence
  microscopic images showing the movement of an actin filament in the
  three dimensional actin-HMM network with $\ca= 9.5~\mu$M ($\xi \sim
  0.6~\mu$m) and $\rah=12$. Note that the leading end of the filament
  is moving out of focus.  \textbf{(a)} The time sequence shows a
  reporter filament of length $16.5$~$\mu$m (see arrow), which within
  $8$~s moves over a distance of $16.5$~$\mu$m at an average speed of
  2~$\mu$m/s.  \textbf{(b)} Superposition of 69 traced contours.
  Traces corresponding to the reporter filament before ($t=0$~s) and
  after ($t=7.8$~s) active transport are shown explicitly.
  \textbf{(c)} Time evolution of the filament position along its average
  contour. Note the sudden onset and halt.}
\end{figure}

Our experimental setup allows us to simultaneously measure the
amplitude of embedded beads and the dynamics of fluorescently
labelled reporter filaments. At the beginning of the experiment,
while the network is still not fully polymerized, these filaments
show a rather uniform diffusive behavior similar to free filaments
in aqueous solutions. Towards the end of the polymerization
process, the movement of the reporter filaments becomes more and
more restricted by the surrounding network.  In this phase, $t
\leq t_\text{sg}$, the actin filaments show thermal undulations
and perform Brownian motion along the cylindrical cages formed by
the surrounding F-actin. In contrast to an actin network
containing myosin~II minifilaments~\cite{humphrey-etal:02}, here
long term directed motion of F-actin is observed only rarely.
During the sol-gel transition, $t_\text{sg} \leq t \leq
t_\text{g}$, any drift that sometimes was observed in the sample
chamber came to a rather abrupt end indicating the stiffening of
the network. At the same time single filaments that shortly before
had hardly shown any fluctuations, suddenly started long term
directed motion inside the three dimensional network (Fig.\
\ref{fig:BetterSequence}\textbf{\mbox{a--c}}).

For our study, $132$ moving filaments were evaluated.  The
phenomenon of moving filaments in the actin-HMM network occurs in
a narrow time slot of some minutes in the proximity of
$t_\text{sg}$, i.e. when ATP depletion starts.  There are also
some rare events of directed filament motion outside this time
window. In some regions of the sample up to $40$\% of the
filaments in focus move, in others only a small fraction is
mobile.  The described phenomena occur under different buffer
conditions, different actin and HMM concentrations and different
molar actin-to-HMM ratios. Also, aging of HMM stock solutions is
affecting the motion.  Therefore, only experiments performed on
the same day with the same actin preparation and HMM stock
solution were compared.  The values obtained for the filament mean
run length and mean velocity are compatible with two dimensional
motility assays at low ATP concentration~\cite{howard_book}. The
longitudinal motion occurred at speeds up to $2.6$~$\mu$m/s over
distances up to $37$~$\mu$m.  Single filaments were moving
continuously for up to $100$~s.  The movement consists of an
acceleration phase, a phase of constant speed and a deceleration
phase (cf. Fig.\ \ref{fig:BetterSequence}\textbf{c}). Due to
experimental limitations, not all the observed filaments could be
monitored in all three phases. In many cases, the filaments moved
a distance of approximately their own visible length and then
paused.

\begin{figure}[bth!]
  \includegraphics[width=\columnwidth]{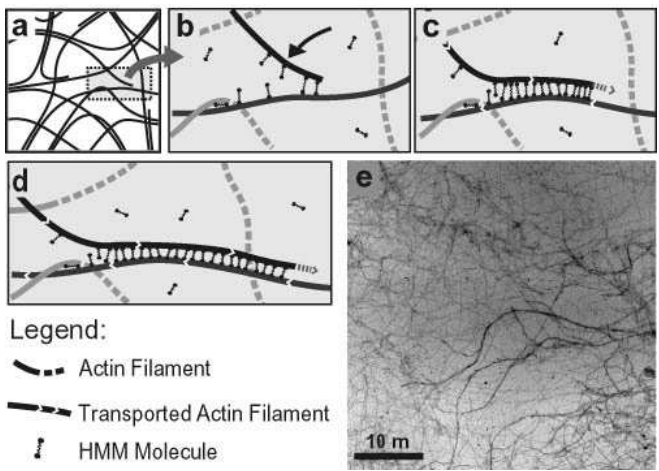}
  \caption{\textbf{(a-d)} Illustration of the ``zipping'' mechanism:
    After one myosin dimer happens to bind two anti-parallel filaments
    simultaneously, thermal fluctuations in the vicinity of the
    binding site are greatly reduced. This enhances the probability
    that nearby myosin molecules also bind to both filaments and leads
    to a nucleation process which zips the filaments together and
    generates relative filament motion.  \textbf{(e)} Electron
    micrograph of the partially bundled network in the vicinity of the
    sol-gel-transition.}
\label{fig:zipping}
\end{figure}

What is causing this sudden onset of directed filament motion in the
vicinity of the sol-gel transition? We interpret it as follows. In a
three-dimensional actin network double-headed myosin can exert
relative forces between filaments only if its heads are simultaneously
attached to two different filaments while one of its heads is
performing a power stroke. Since under saturating ATP condition
myosin~II has a very low duty ratio, such events are extremely rare.
Actin filaments can be transported only if a certain amount of motors
work together like in the thick filament of a sarcomere. A similar
effect seems to be at work in recent experiments where myosin II
molecules arrange themselves into
minifilaments~\cite{humphrey-etal:02}. In the present study we use
HMM, which does not form such minifilaments.  Hence cooperativity of
the motors must be achieved by a different route. Now, ATP is needed
to trigger the dissociation of the motor from the filament, and an
initial ATP concentration is depleted due to the ATP-hydrolysis of the
motor. This prolongs the time a myosin head stays bound to a filament
(duty ratio), which increases the chance for the other head to also
bind to a filament while the first one is still bound.  As a
consequence one can have two opposing effects.  First, an increasing
fraction of HMM molecules acts as cross-linkers in the actin network,
and stiffens it. This explains the sol-gel transition observed by the
sudden decrease in the amplitude of the oscillating magnetic bead (see
Fig.~\ref{fig:SGT}), and also the stop of a drift in the sample
chamber as mentioned above.  Second, the probability that myosin
molecules can exert directed forces onto actin filaments is increased.
One may envisage the following ``zipping'' mechanism, illustrated in
Fig.\ \ref{fig:zipping}\textbf{\mbox{a--d}}. After one myosin dimer
happens to bind two anti-parallel filaments simultaneously, thermal
fluctuations in the vicinity of the binding site are reduced.  This
enhances the probability that nearby myosin molecules also bind to
both filaments and leads to a nucleation process which zips the
filaments together and generates relative filament motion. During the
transport the overlap region may grow.  This would explain the sudden
onset of filament motion with a moderate acceleration phase.  Such an
interpretation is also supported by the broad distribution of run
lengths and run times seen in our traces; zipping of the filaments can
either occur at one of the ends leading to a maximal run length
identical to the filament length or along the contour which
potentially greatly reduces the run length due to entanglement
effects.

We have explored how the observed effects depend on actin, $\ca$, and
HMM concentration, $\cm$, and obtained the following tendencies.
Doubling $\ca$ from $9.5$ to $19 \, \mu$M while keeping $\cm=0.38 \,
\mu$M constant reduces the average transport length by a factor of
$3$.  Within the ``zipping'' mechanism this can be attributed to
entanglement effects.  Reducing $\cm$ from $1.05$ to $0.38$~$\mu$M
with $\ca = 9.5 \, \mu$M fixed lowers the average transport speed from
$0.5$ to $0.31 \, \mu$m/s. This may be understood as a result of the
reduced number of motors in the zipped region between the filaments.
In general we observe broad distributions in running length, time and
speed. This may be due to heterogeneities in the network (see Fig.\
\ref{fig:zipping}\textbf{e}), length distribution of actin filaments,
their relative orientation \cite{kruse-julicher:00} as well as the
location of the initial seed in the zipping process.  Future
experiments with a much larger number of traced filaments and samples
are certainly necessary to explore the details of this fascinating
dynamic phenomenon.

We thank H. Kirpal for HMM preparations, M. Rusp for actin
preparations and her help with the electron microscope, and J.
Schilling for providing the image processing software ``Open Box''.
This research has been financially supported by the Deutsche
Forschungsgemeinschaft (SFB 413) and the Fonds der Chemischen
Industrie. A.P. is supported by a Marie-Curie Fellowship under
contract no. HPMF-CT-2002-01529. We acknowledge helpful discussions
with K. Kroy and G. Lattanzi.

\end{document}